\documentclass[a4paper]{mem}
\usepackage{natbib}
\usepackage{graphicx}
\usepackage[a4paper]{hyperref}
\idline{73}{1}

\begin{document}

\title{Geminga: a tale of two tails, and much more}

   \author{P.A. Caraveo\inst{1}, G.F. Bignami \inst{2,3}, A. De Luca \inst{1,4}, A. Pellizzoni\inst{1}, S. Mereghetti\inst{1},  R.P. Mignani\inst{5}, A. Tur\inst{2}, W. Becker\inst{6}}

   \offprints{P.A. Caraveo}

   \institute{INAF/IASF ``G. Occhialini'', Via Bassini 15, 20133 Milano, Italy \email{pat@mi.iasf.cnr.it}
              \and Centre d'Etude Spatiale des Rayonnements, CNRS-UPS, 9, Avenue du Colonel Roche, 31028, Toulouse Cedex 4, France 
              \and Universit\`a degli Studi di Pavia, Dipartimento di Fisica Nucleare e Teorica, Via Bassi 6, 27100 Pavia, Italy
              \and Universit\`a di Milano Bicocca, Dipartimento di Fisica, P.za della Scienza 3, 20126 Milano, Italy
              \and ESO, Karl Schwarzschild Strasse 2, D-85740, Garching bei M\"unchen, Germany
              \and Max-Plank Institut f\"ur Extraterrestrische Physik, 85741 Garching bei M\"unchen, Germany
          }

   \abstract{We report on the deep (100 ksec) XMM-Newton/EPIC observation of the field of the Geminga pulsar.
   The unprecedented throughput of the instrument allowed to detect  two elongated parallel x-ray tails trailing the neutron
star. They are aligned with the object's supersonic motion, extend for $\sim$2', and have a nonthermal spectrum produced by electron-synchrotron emission
in the bow shock between the pulsar wind and the surrounding medium. Electron lifetime against synchrotron cooling matches the source transit time over
the x-ray features' length. Such an x-ray detection of a pulsar bow shock allows us to gauge the pulsar electron injection
energy and the shock magnetic field while constraining the angle of Geminga's motion and the local matter density. 
We give also preliminary results on the timing and spectral analysis of the $\sim$63,000 photons collected from the neutron star.

   }
   \authorrunning{P.A. Caraveo et al. et al.}
   \titlerunning{Geminga: the tails, and more}
   \maketitle

\section{Introduction}

   \begin{figure*}[htb!]
   \centering
   \resizebox{\hsize}{!}{\includegraphics[clip=true]{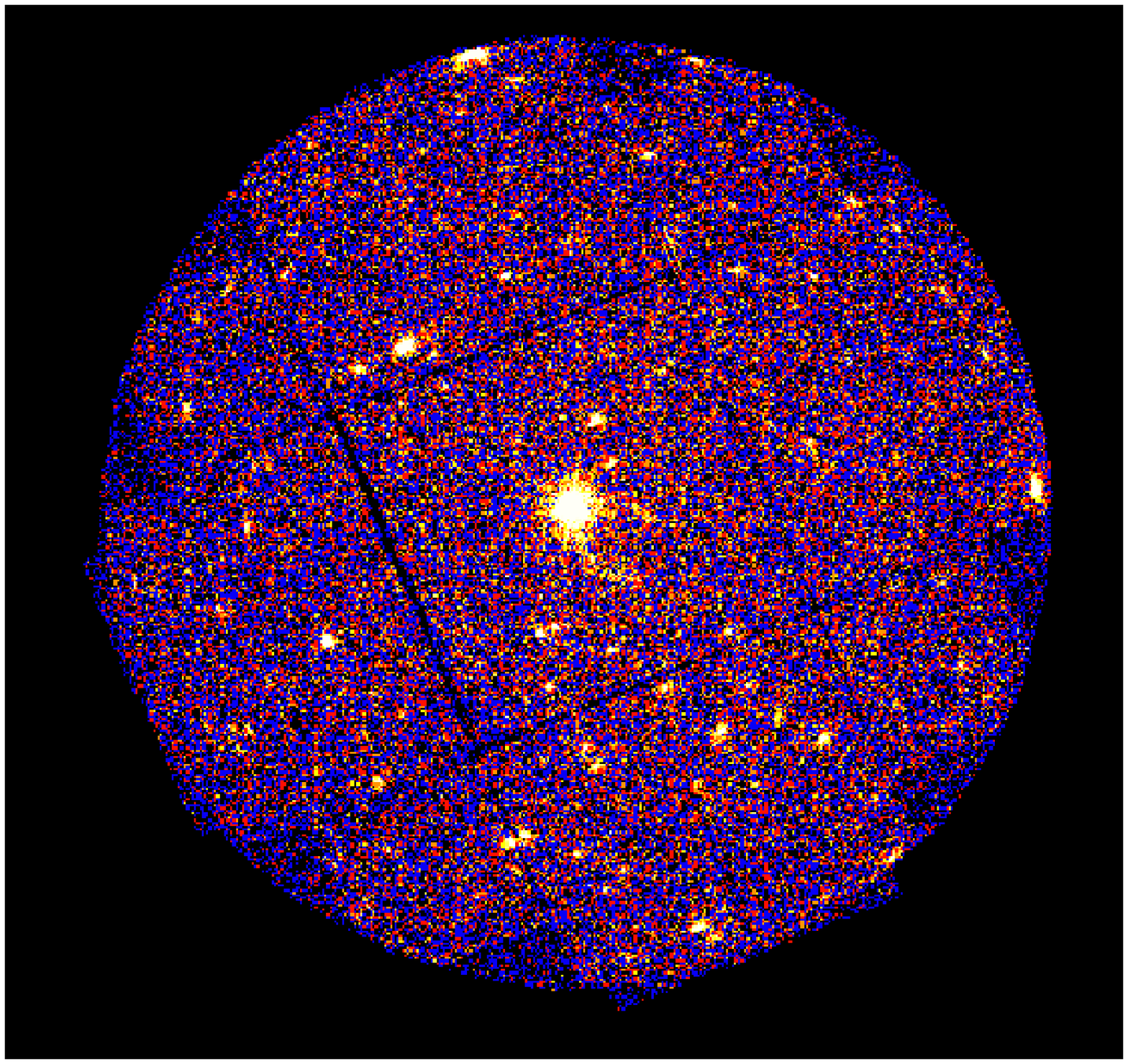}
   \includegraphics[clip=true]{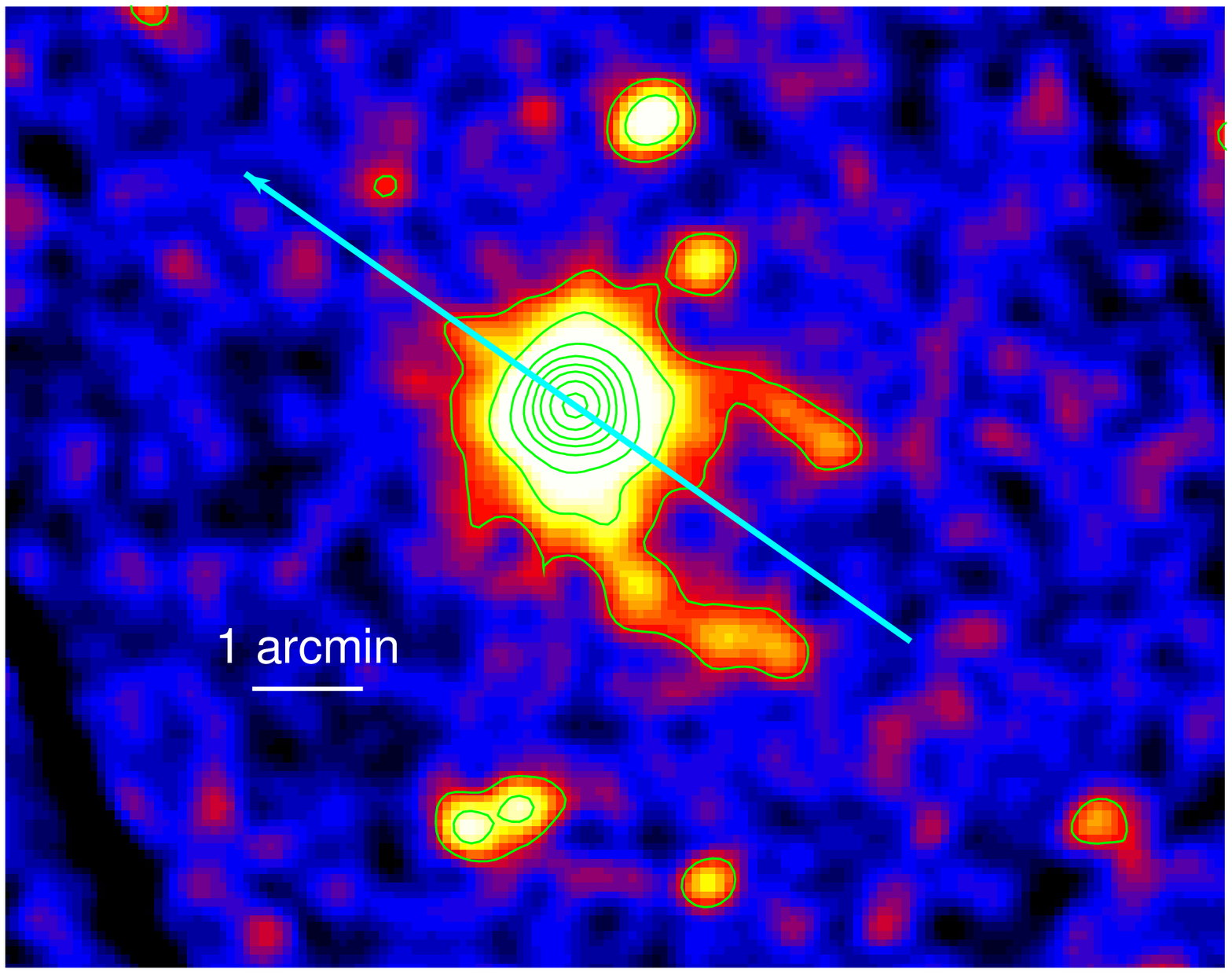}}
     \caption{(left) The XMM-Newton view of the field of Geminga. Data from 
     the MOS1 and MOS2 cameras have been merged to produce the image. Events in the 0.3-5 keV range have been selected.
     The exposure time is of $\sim77$ ksec per camera. Geminga is the bright source close to the center of the image; 
     the tails can be seen as two faint diffuse emission patterns emerging from the source. 
     Many ($\sim100$) serendipitous sources are also visible 
     in the field.
     (right) Inner part of the field, shown after gaussian smoothing. The emission from Geminga outshines the tails 
     up to $\sim 40''$ from the source.
     The tails are $\sim$2 arcmin long and cover an area
     of $\sim$2 square arcmin. They show a remarkable symmetry with respect to the pulsar proper motion direction, 
     marked by the arrow.
               }  
        \label{fit1}
    \end{figure*}

Geminga is a nearby, middle-aged isolated neutron star (Bignami \& Caraveo, 
1996). Proximity is  a key-parameter for understanding the multiwavelength 
behaviour of  this source discovered in high-energy gamma-rays  and 
later identified in X-rays (Bignami et al., 1983) and optical wavelengths (Bignami 
et al., 1988). The smoking gun, confirming the previous work based on the 
interpretation of positional coincidence, came with the ROSAT discovery of 
a 237 msec periodicity (Halpern \& Holt, 1992), immediately seen also at higher energy in 
contemporary EGRET data (Bertsch et al. 1992) as well as in COS-B archival data 
(Bignami \& Caraveo, 1992). At the same time, significant proper motion was discovered at 
optical wavelengths (Bignami et al., 1993), definitely linking the proposed 
counterpart to a fast-moving, pulsar-like object, the distance to which 
was later nailed down to 160 pc through its optical parallax (Caraveo et al., 1996).   
Thus, Geminga qualifies as a pulsar, and as such is listed in the pulsar 
catalogue\footnote{http://www.atnf.csiro.au/research/pulsar/psrcat/}, although it has not been detected at radio 
wavelengths.  Indeed, the absence of a radio signal prompted Caraveo, 
et al. (1996b) to use this source as the prototype of a new 
class of pulsars: the radio-quiet ones, now encompassing  the so called ``dim 
sources'' (Treves et al., 2000), the CCO (Pavlov et al., 2002),  the next Geminga   
(Halpern et al., 2002) and other potential candidates, including a number of proposed 
counterparts of EGRET sources. 
Gamma-ray emission is not a common characteristics of these radio quiet 
INSs. Rather, their common denominator is their  thermal soft X-ray emission  
coupled with very faint (if any) optical radiation.
Here we report on the XMM-Newton/EPIC data of a long  Geminga exposure which almost 
triples the number of soft photons available from this source.
Such improved statistics allows to study the object's light curve as a function 
of the photon energy.
The total (average) source spectral shape can be assessed using three 
independent instruments, while the time-tagging of photons allows, for the first 
time,  to follow the evolution of the source spectral shape as a function of  
phase.

\section{The EPIC data}

   \begin{figure*}[htb!]
   \centering
  \includegraphics[angle=-90,width=11cm]{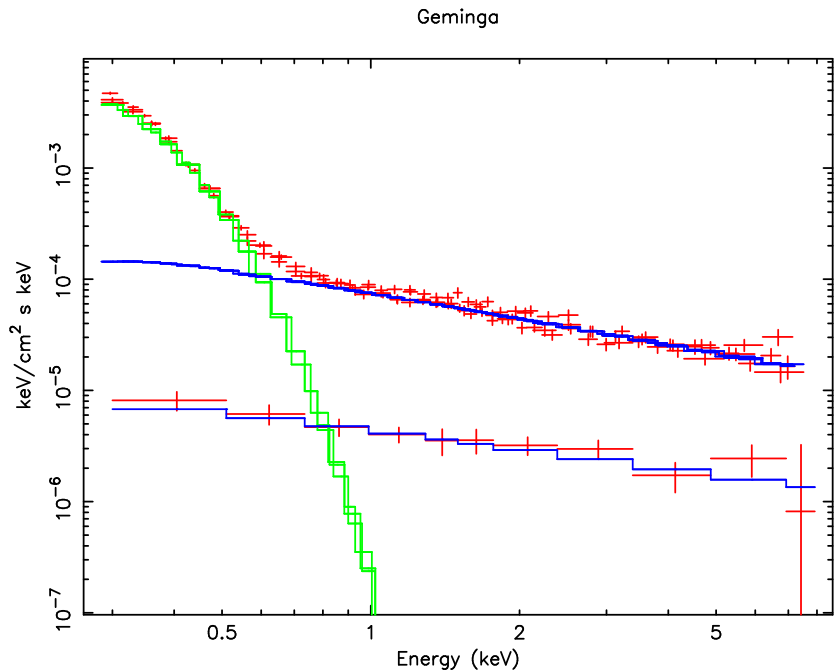}
   \caption{Unfolded spectrum of Geminga (upper curve) and its tails (lower curve).
   The spectrum of the pulsar (pn, MOS1 and MOS2 data points are plotted together)
   is well reproduced by the sum of a blackbody function and a hard power law (see text).
   The spectrum of the tails (the sum of MOS1 and MOS2 data is plotted) is also well 
   described by a similar hard power-law curve.}
    \end{figure*}

XMM-Newton performed  a 100 ksec exposure on Geminga on April 4th,  
2002. 
 While the two MOS EPIC cameras were operated with the medium filter in their
``full frame'' mode (Turner et al. 2001), the pn camera was operated with the 
thin filter in ``small window'' mode to allow for  accurate timing of  
source photons (Str\"uder et al., 2001).  Data were processed with the 
XMM Newton Science Analysis Software (SAS version 5.4.1). After removing 
time intervals with high particle background and correcting for the dead 
time, we obtain a net exposure time of  55.0 ksec for the pn camera and 
76.9 and 77.4 ksec for the MOS1 and MOS2, respectively. 
The Geminga count rates (0.2$<$E$<$7 keV) are  0.807$\pm$0.004 counts s$^{-1}$ for the 
pn and 0.119$\pm$0.001 counts s$^{-1}$  and   0.123$\pm$0.001 counts s$^{-1}$ for the MOS1 and MOS2, 
respectively.
The sum of the MOS1 and MOS2 images is shown in Figure 1.  Besides Geminga, which shines at the center of the 
image,  more than  100 serendipitous sources have been detected. Identification work is in progress 
on such sources and will be reported elsewhere.
For a thorough  discussion of the ``tails'', which are seen by EPIC for the 
first time as trailing Geminga and well aligned with the source proper motion, 
the reader is referred to Caraveo et al. (2003). Briefly, we recall here the most important 
characteristics of this newly-discovered X-ray feature of Geminga.

The tails are two patterns of diffuse emission, originating close to 
Geminga (they cannot be resolved in proximity of the bright point source
closer than 40$''$ at most)
and extending up to $\sim3$ arcmin away from the pulsar, well aligned with the NS proper motion, 
with a thickness of $\sim~20''-30''$.
Their spectrum is well reproduced by a slightly absorbed ($N_H \sim 
10^{20}~cm^{-2}$) power law with photon index $\Gamma \sim 1.6$. At the pulsar
distance (160 parsecs), their unabsorbed 0.3-5 keV flux corresponds to a 
luminosity of 6.5$\times 10^{28}$ erg s$^{-1}$ 
($\sim2 \times 10^{-6}$ of Geminga's rotational energy loss).  
Possible contributions from point sources are estimated to be negligible.

The shape of the tails is reminiscent of the projection on the plane of
the sky of an empty paraboloid of X-ray emission, the edges of which
show up brighter because of a limb effect. Such a morphology is naturally 
explained in terms of a bow-shock formed between the pulsar relativistic wind 
and the dynamical pressure generated by its supersonic motion through the 
interstellar medium.

The hard, power-law spectrum of the tails can be  explained by synchrotron 
emission of high-energy electrons accelerated by the pulsar, girating in the 
shocked interstellar magnetic field. 
Such an interpretation provides a direct gauge 
of the pulsar wind injection energy, demonstrating that Geminga accelerates 
electrons up to 10$^{14}$ eV, a value very close to the upper limits expected 
on the basis of the pulsar's energetics, allowing also to constrain the 
local interstellar magnetic field in the range 2-3 $\mu G$.
The Larmor radius of the emitting electrons in the bow-shock magnetic field
($\sim~27''$ at the distance of Geminga) is found to be  
in excellent agreement with the observed thickness of the tails. 
Moreover, electron lifetime against synchrotron emission 
($\sim~800$ years) matches the pulsar transit time over the 
X-ray features' length ($\sim~1000$ years, on the basis of the well known 
proper motion of Geminga), supplying a final, independent consistency check to 
our model.

The observed geometry of the tails, compared to a 3-D bow-shock model assuming 
a spherical pulsar wind in an homogeneous ISM, allows also to constrain 
the angle of the source motion with respect to the plane of the sky to be less 
than $\sim30^{\circ}$, and therefore to assess its 3-D space velocity.
The detection of the pulsar bow-shock represents also an important  way to probe 
the interstellar medium, constraining its density to be in the range 0.06-0.15 
atoms cm$^{-3}$, in good agreement with
the expected value of the ISM density for the region around Geminga (Gehrels \&
Chen 1993).

   \begin{figure*}[htb!]
   \centering
  \includegraphics[angle=-90,width=11cm]{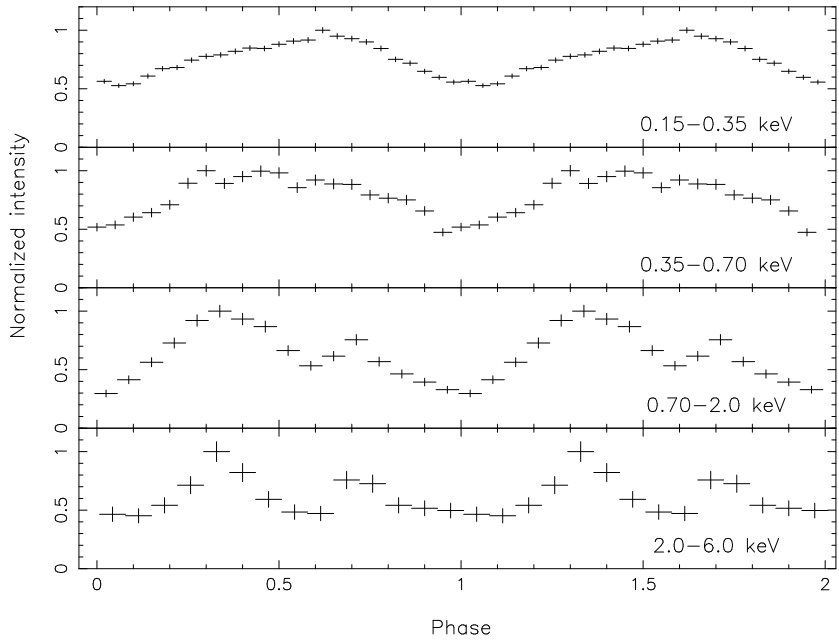}
   \caption{Folded light curves of Geminga in different energy ranges.
   Morphology variations of the pulse profile as a function of energy are
   clearly evident (see text). }
    \end{figure*}


The good EPIC statistics allows for an accurate spectrum to be drawn, 
for the first time, of both the pulsar and its tails, in the range 0.2 to $\sim$7 keV. 
Fig.2 shows such time-averaged spectrum of both the pulsar (upper curve) and its surrounding 
diffuse emission, drawn on the source energy flux scale. Note that the 
upper curve plots both the pn and the 2 MOS data (superimposed and undistinguishable),
while the lower curve for the extended emission (tails) contains MOS data only,
since the pn camera was used in ``small window'' mode.
While a more detailed interpretation of the source physics as portrayed by 
Fig.2 will be the topic of an upcoming work,
we propose here a few qualitative, obvious comments. The point-source 
spectrum (upper curve) shows with a striking clarity that two different
mechanisms are at work. Below $\sim0.7$ keV, the emission is undoubtely
thermal, well fit by a temperature of $\sim$43 eV, implying an emitting surface 
of $\sim9$ km at 160 parsec.
For the whole energy decade 0.7-7 keV, Geminga's spectrum is equally well
fit by a single power law, with photon index $\alpha \sim$1.85.
The simplest interpretation of such hard power law (already suggested
by the ROSAT data of Halpern \& Holt, 1992) is that of synchrotron emission by energetic 
electrons radiating in the pulsar magnetic field.

For the diffuse emission from the tails (lower curve of Fig.2) a striking similarity 
is apparent with the hard spectral shape of the point source.
This points to a similar physical origin between the source and the tails hard X-ray 
photons. The latter, however, must necessarily be created by the high energy
(10$^{14}$ eV) end of the pulsar electron spectrum, since they radiate in the compressed 
IS magnetic field of $\sim10^{-5}$ G.
 

\section{Timing analysis}
 
Exploiting the much improved EPIC energy resolution with respect to both ASCA and ROSAT,
our data allow, for the first time, to render apparent the pulsar light curve morphology 
variation as a function of photon energy.


Fig.~3 shows such energy-resolved time curves for four different energy ranges: 0.15-0.35, 
0.35-0.7, 0.7-2.0 and 2.0-6.0 keV. The light curve morphology variation is apparent. 
The lower energy, thermal emission has a smooth light curve featuring a single, wide peak.
At high energy, two peaks are present during each phase interval, separated by $\sim$0.4 phase.
It is, at the moment, not possible to relate in absolute phase the EPIC X-ray peaks to the EGRET
$\gamma$-ray peaks; however, we note the striking similarity between the two non-thermal
light curves of the same object.
A phase shift of about 0.3 phase is also obviously present between the single peak of the thermal emission
and the highest of the two non-thermal peaks.
As expected, the ``intermediate'' energy range (0.35 to 0.7 keV) shows a somewhat mixed
behaviour due to the symultaneous presence of the two components. Such components are
obviously of very different physical origin, also possibly in terms of production zone.
Finally, we note that the pulsed fraction of the radiation as a function of energy 
does not seem to show the spectacular variation proposed by Jackson et al. (2002) on the
basis of ASCA data. In particular, the EPIC data do not confirm a pulsed fraction as high as,
e.g., 80\% or greater at high energy  (E$>$4 keV).
A final analysis of all aspects of light curve morphology variation with energy is currently
in progress and will be published elsewhere.

\section{An EPIC remark}
 
The EPIC observation yielded a total of  63,000 photons. 
The vast majority of  the EPIC harvest is due to the pn 
detector with  44,400 photon with the MOS1 and MOS2 detectors contributing 
9,150 and  9520 photons respectively. The tails, which are made up by  few 
hundreds photons, are not included in this budget.
When compared with the 27,000 soft X-ray photons gathered during the ROSAT 
lifetime in the energy range (0.1-2.4 keV)  and the 6,500 collected by 
ASCA in the energy range (0.7-8 keV), our EPIC observations stands out for 
its much improved statistics. Since the part of the spectrum benifitting most 
from the high troughput of the XMM-Newton telescope is the ``high energy'' 
one, it comes as no surprise that the EPIC spectrum constrain much 
better the high energy part of the source spectrum as well as its time-resolved behaviour.

\bibliographystyle{aa}

\end{document}